\documentclass[letterpaper]{article}
\usepackage[utf8]{inputenc}
\usepackage[tmargin=1in,bmargin=1in,lmargin=1.0in,rmargin=1in]{geometry}
\usepackage{hyperref}
\usepackage{amsthm, amsmath, amsfonts, bbm}
\title{The Prophet Inequality Can Be Solved Optimally with a Single Set of Samples}
\small{
\date{August 25, 2018}
}

\newtheorem{lemma}{Lemma}
\newtheorem{corollary}{Corollary}

\begin{document}
\setcounter{page}{0}

\begin{titlepage}
\author{Jack Wang \thanks{Harvard University, jackwang@college.harvard.edu}}
\maketitle

\begin{abstract}
The setting of the classic prophet inequality is as follows: a gambler is shown the probability distributions of $n$ independent, non-negative random variables with finite expectations. In their indexed order, a value is drawn from each distribution, and after every draw the gambler may choose to accept the value and end the game, or discard the value permanently and continue the game. What is the best performance that the gambler can achieve in comparison to a prophet who can always choose the highest value? Krengel, Sucheston, and Garling solved this problem in 1978, showing that there exists a strategy for which the gambler can achieve half as much reward as the prophet in expectation. Furthermore, this result is tight.

In this work, we consider a setting in which the gambler is allowed much less information. Suppose that the gambler can only take one sample from each of the distributions before playing the game, instead of knowing the full distributions. We provide a simple and intuitive algorithm that recovers the original approximation of $\frac{1}{2}$. Our algorithm works against even an \textit{almighty adversary} who always chooses a worst-case ordering, rather than the standard \textit{offline adversary}. The result also has implications for mechanism design -- there is much interest in designing competitive auctions with a finite number of samples from value distributions rather than full distributional knowledge.

\end{abstract}
\renewcommand{\thepage}{}

\end{titlepage}

\section{Introduction}
Consider the following probabilistic game. Let $X_1,...,X_n$ be a sequence of independent random variables supported on $\mathbb{R}^+$ with $\mathbb{E}[\max_iX_i] \leq \infty$. A gambler is shown all of the distributions and the game proceeds by drawing values one at a time from the distributions in the order that they are indexed. At any point the gambler may choose to either take the value and end the game, or discard the value irrevocably and observe the next one. Krengel, Sucheston and Garling \cite{KS78} showed that there exists a stopping rule $\tau$ such that
\begin{equation}
2\mathbb{E}[X_\tau] \geq \mathbb{E}\big{[}\max_i X_i \big{]}
\end{equation}
In other words, there exists a strategy for the gambler to achieve half as much value in expectation as the prophet who is always able to select the maximum value. Furthermore, this approximation factor is tight. Consider the following example with two random variables, first demonstrated in \cite{KS78}. Let $X_1$ be the random variable which always returns a value of 1, and let $X_2$ be the random variable which has value $\frac{1}{\epsilon}$ with probability $\epsilon$ and value 0 otherwise. The gambler can choose the first and achieve a fixed value of 1, or she can choose to gamble on the second distribution and achieve an expected value of 1. However, the prophet's optimal strategy is to choose the second distribution when it achieves the high value, and choose the first otherwise. Thus the prophet's reward is $\epsilon\frac{1}{\epsilon} + (1 -\epsilon)1 = 2-\epsilon$.

In this work, we consider a situation in which the gambler has considerably less power. What if instead of having full knowledge of the distributions, the gambler was only allowed to observe one sample from each of the distributions before playing the game? Could she even achieve any constant-competitive ratio? Surprisingly, we show that the gambler can still achieve a $2$-approximation with this limited information. Our algorithm is simple and intuitive -- we simply set the highest valued sample as a fixed threshold and accept the first value which exceeds the threshold. In addition to being an independently interesting result, this work has practical implications for mechanism design, showing competitive guarantees on welfare and revenue for a prior-independent order-oblivious posted price mechanism for a single item.

\section{Related Work}
Since its introduction in \cite{KS77} and \cite{KS78}, the prophet inequality has spawned a vast line of literature in the areas of online selection problems and mechanism design. It would be impractical to summarize all of the literature here, but we will touch on some parts that are relevant and interesting.

The constraint of being able to choose one item has been expanded to many combinatorial domains including multiple choices \cite{Alaei}, matroids \cite{matroid}, and general down-closed set systems \cite{beyond}. The connection between prophet inequalities and auction design was first noted by Hajiaghayi, Kleinberg, and Sandholm \cite{HKS-aaai}. Chawla et. al \cite{CHMS} explored this in detail, showing that prophet inequalities could be used to answer questions about the performance of posted price mechanisms. Recently, Correa et. al explored the reverse direction, showing that posted price mechanisms and their guarantees could be turned into results about prophet inequalities \cite{correa}. Many threads involving combinatorial prophet inequalities and posted price mechanisms were united and generalized by D\"utting et al. \cite{dutting}. Another recent connection was noted by Lee and Singla \cite{lee} between prophet inequalities and contention resolution schemes, a technique for solving combinatorial optimization problems.

The closest relation to our work is Azar, Kleinberg, and Weinberg's exploration of prophet inequalities with a limited number of samples \cite{azar2}. In that work, they derive an asymptotically optimal result for single sample prophet inequalities in the $k$-choice model, which certainly encompasses a single choice. However, their result does not include a tight constant factor, and in particular does not characterize a scenario in which having single samples is just as powerful as having the full distribution.

\section{Algorithm}
The algorithm proceeds as follows. Let $(X_1,...,X_n)$ be the vector of random variables and $(v_1,....,v_n)$ be the corresponding vector of observed samples. If $h$ is the index of the highest value, we set $v_h$ as a threshold and immediately accept the first value which exceeds the threshold. We assume without loss of generality that there is a strict ordering on all of the values, i.e. no two values are equal to each other. To address the case in which values can be equal, we use a standard smoothing technique, which is explained in section \hyperref[sec:smoothing]{3.4}. We now proceed with the analysis.

\subsection{A Simplifying Simulation}
To analyze the guarantee of this algorithm, we introduce a simulated version of the game. We will argue that this simulation is equivalent to the original game and show that the algorithm guarantees a $2$-approximate solution under this simulation, which then applies to the original game as well.

Suppose that the person running the game draws two sets of values, $(v_1^1,....,v_n^1)$ and $(v_1^2,...,v_n^2)$, storing them before the game begins. Let $C^n \sim Bern(\frac{1}{2})^n$ be a random vector representing $n$ independent coin flips. If $C^n_i = 0$, then they assign $v_i^1$ to arrive in the sample phase and $v_i^2$ to arrive as a real value, and vice versa if $C^n_i=1$. Let $S$ denote the vector of values which were assigned as samples, and $R$ the vector of values assigned as reals. It is easy to see that $S$ and $R$ are distributed exactly as what the gambler would see in the original game -- the distribution of the drawn values is unaffected by conditioning on the coin flip, as all the values are drawn independently from the original distributions.

We would like to argue that over the randomness of $C^n$, given fixed $(v_1^1,....,v_n^1)$ and $(v_1^2,...,v_n^2)$, the gambler will in expectation receive a $2$-approximate value (in comparison to the prophet who can always pick the maximum value that was sorted into the real values). In other words, we want to show

\begin{equation}\label{eq:2}
 \mathbb{E}_{C^n}[\text{ALG}| v_1^1,v_1^2,...,v_n^1,v_n^2] \geq \frac{1}{2} \mathbb{E}_{C^n}\Big[\max_{ v_i^k \in R} v_i^k| v_1^1,v_1^2,...,v_n^1,v_n^2\Big]
\end{equation}

If we have shown \hyperref[eq:2]{(2)}, then we can take the expectation over the draws to obtain
\begin{equation}
 \mathbb{E}_{X_1,...,X_n}\big[\mathbb{E}_{C^n}[\text{ALG}| v_1^1,v_1^2,...,v_n^1,v_n^2]\big] 
\geq \frac{1}{2}\mathbb{E}_{X_1,...,X_n}\Big[ \mathbb{E}_{C^n}\Big[\max_{ v_i^k \in R} v_i^k| v_1^1,v_1^2,...,v_n^1,v_n^2\Big]\Big] 
\end{equation}

\begin{equation}
\mathbb{E}_{X_1,...,X_n,C^n}[\text{ALG}] \geq \frac{1}{2}\mathbb{E}_{X_1,...,X_n,C^n}\Big[\max_{ v_i^k \in R} v_i^k\Big]
\end{equation}
This means that under our simulated game, the algorithm achieves a 2-approximation. Since the random vector $C^n$ simply decides which of two draws from the same distribution arrive as samples or reals, no matter if $C^n$ is fixed or random, we still have $S$ and $R$ distributed independently as $ (X_1,...,X_n)$. So the above also implies
\begin{equation}
\mathbb{E}_{X_1,...,X_n}[\text{ALG}| C^n = (0,...,0)] \geq \frac{1}{2}\mathbb{E}_{X_1,...,X_n}\Big[\max_{ v_i^k \in R} v_i^k | C^n = (0,...,0)\Big]
\end{equation}
This is precisely the setting of the original game, as values are picked into $S$ and $R$ without any permutation. Thus, it only remains to prove \hyperref[eq:2]{(2)} to show that our algorithm achieves a 2-approximation in the original game.

\subsection{Expected Maximum}\label{sec:3.2}
Suppose that the person running the game has already drawn  $(v_1^1,....,v_n^1)$ and $(v_1^2,...,v_n^2)$ but has yet to sort them into $S$ and $R$. Let $Y_i^k$ be the probability that given a fixed set of draws, $v_i^k$ becomes the maximum real value over the randomness of $C^n$. Then by linearity of expectation the expected maximum conditioned on these draws can be calculated as
\begin{equation}
\sum_{i,k} Y_i^k v_i^k
\end{equation}

Suppose we order all of the draws descending by weight, giving us $(v_{i_{1}}^{k_{1}},...,v_{i_{2n}}^{k_{2n}})$. Scanning through the ordering from first to last, at some point we will see some $v_{i_m}^{k_m}, v_{i_{m'}}^{k_{m'}}$ such that $i_{m} = i_{m'}$. In other words, we will encounter two values from the same distribution for the first time. Let this first value which caused us to record a repeated distribution be $v_{\text{repeat}}$, and let the other value which came from the same distribution be $v_{\text{repeat-pair}}$. To pause for clarity, we should understand that $v_{\text{repeat-pair}} > v_{\text{repeat}}$, or that $v_{\text{repeat-pair}}$ is the earlier of the two. 

Let $T$ be the set of all values greater than $v_{\text{repeat}}$. Only the values in $T$ and $v_{\text{repeat}}$ have a non-zero probability of being the maximum value in $R$. This is because one of $v_{\text{repeat}}$ and $v_{\text{repeat-pair}}$ must become sorted into $R$, so any value that is smaller than both $v_{\text{repeat}}$ and $v_{\text{repeat-pair}}$ will always have a larger value sorted onto the same side as it.

Now we can calculate the value of each $Y_i^k$ for $v_{i}^k \in T$. What would need to happen for $v_i^k$ to become the maximum real value? First, it would obviously need to be sorted into $R$. Then, every value in $T$ that is larger than it must be sorted into $S$. Let $T_i^k$ denote the set of values in $T$ which are greater than $v_i^k$. Thus we conclude that for $v_i^k \in T$,
\begin{equation}
Y_i^{k} = \bigg{(}\frac{1}{2}\bigg{)}^{|T_i^k| + 1}
\end{equation}

Suppose $v_i^k = v_{\text{repeat}}$, then it needs to be sorted into $R$, but this is the same event that $v_{\text{repeat-pair}}$, an element in $T$, is sorted into $S$. All of the other values in $T$ must be samples as well. Therefore if $v_i^k = v_{\text{repeat}}$
\begin{equation}
Y_i^k = \bigg{(}\frac{1}{2}\bigg{)}^{|T|}
\end{equation}
This gives us a clearer expression of the expected maximum over the randomness of $C^n$ as
\begin{equation}\label{eq:9}
\bigg(\sum_{v_i^k \in T} \bigg{(}\frac{1}{2}\bigg{)}^{|T_i^k| + 1} v_i^k \bigg) + \bigg{(}\frac{1}{2}\bigg{)}^{|T|}v_{\text{repeat}}
\end{equation}

\subsection{Expected Value of Algorithm}

Next we want to calculate a lower bound on the expected value of the algorithm. Certainly it would be a worst case scenario if of all the values in $R$ which exceed the threshold, the algorithm always selects the smallest one. Let $Z_i^k$ be the probability that $v_i^k$ is selected under this assumption. Only values in the set $T$ will ever be selected by the algorithm. This is because one of $v_{\text{repeat}}$ or $v_{\text{repeat-pair}}$ must be sorted into $S$, so the threshold is no lower than $v_{\text{repeat}}$.

For a certain value $v_i^{k} \in T$, what needs to happen for it to be selected under our assumption?
\begin{lemma}\label{lemma:1}
If the lowest threshold-exceeding value is always selected by the algorithm, a value $v_i^k$ is selected if and only if the highest value in $S\cup R$ which is lower than $v_i^{k}$ is the maximum in $S$.
\end{lemma}

\noindent\textit{Proof.} Suppose that the highest value lower than $v_i^k$, which we denote by $v_{\text{threshold}(i,k)}$, is the maximum value in $S$. Then $v_i^k$ must be in $R$, as otherwise it would have been the maximum in $S$. $v_i^k$ will be selected as the smallest threshold-exceeding value in $R$ is selected, and there are no values in between $v_{\text{threshold}}$ and $v_i^k$. Suppose that $v_{\text{threshold}(i,k)}$ is not the maximum in $S$. If the maximum in $S$ is smaller, that means that $v_{\text{threshold}(i,k)}$  is not in $S$ and therefore in $R$, and it exceeds the threshold, so it or some even smaller value will be selected over $v_i^k$. If the threshold is larger, then the threshold is greater than or equal to $v_i^k$, so $v_i^k$ will not be selected.

Now for all values, is clear due to symmetry that the probability that a value is the maximum in $R$ is the same probability that it is the maximum in $S$. Now we can re-use our insight from \hyperref[sec:3.2]{3.2} combined with Lemma \hyperref[lemma:1]{1} to clearly calculate the probability that each value is selected by the algorithm over the randomness of $C^n$. Let $v_{\text{last-in-T}}$ be the element in $T$ with the lowest weight. Recall that $T_i^k$ denotes the set of elements in $T$ larger than $v_i^k$. Then for all $v_i^k \in T - \{v_{\text{last-in-T}}\}$,
\begin{equation}
Z_i^k = \bigg{(}\frac{1}{2}\bigg{)}^{|T_i^k| + 2}
\end{equation}

For $v_i^k = v_{\text{last-in-T}}$, 
\begin{equation}
Z_i^k = \bigg{(}\frac{1}{2}\bigg{)}^{|T|}
\end{equation}
This lets us conclude that the expected value of the selected element is at least
\begin{equation}
\bigg(\sum_{v_i^k \in T - \{v_{\text{last-in-T}}\}} \bigg{(}\frac{1}{2}\bigg{)}^{|T_i^k| + 2} v_i^k \bigg) + \bigg{(}\frac{1}{2}\bigg{)}^{|T|}v_{\text{last-in-T}}
\end{equation}
\begin{equation}
 = \frac{1}{2} \Bigg[ \bigg(\sum_{v_i^k \in T - \{v_{\text{last-in-T}}\}} \bigg{(}\frac{1}{2}\bigg{)}^{|T_i^k| + 1} v_i^k \bigg)+ \bigg{(}\frac{1}{2}\bigg{)}^{|T| -1}v_{\text{last-in-T}}\Bigg]
\end{equation}
\begin{equation}
 = \frac{1}{2} \Bigg[ \bigg(\sum_{v_i^k \in T - \{v_{\text{last-in-T}}\}} \bigg{(}\frac{1}{2}\bigg{)}^{|T_i^k| + 1} v_i^k\bigg) + \bigg{(}\frac{1}{2}\bigg{)}^{|T|}v_{\text{last-in-T}} +  \bigg{(}\frac{1}{2}\bigg{)}^{|T|}v_{\text{last-in-T}}\Bigg]
\end{equation}
\begin{equation}
 = \frac{1}{2} \Bigg[ \bigg(\sum_{v_i^k \in T} \bigg{(}\frac{1}{2}\bigg{)}^{|T_i^k| + 1} v_i^k \bigg)+ \bigg{(}\frac{1}{2}\bigg{)}^{|T|}v_{\text{last-in-T}}\Bigg]
\end{equation}
\begin{equation}
\geq \frac{1}{2}\Bigg{[} \bigg( \sum_{v_i^k \in T} \bigg{(}\frac{1}{2}\bigg{)}^{|T_i^k| + 1} v_i^k \bigg) + \bigg{(}\frac{1}{2}\bigg{)}^{|T|}v_{\text{repeat}}\Bigg{]}
\end{equation}
\begin{equation}
=\frac{1}{2} \mathbb{E}_{C^n}\Big[\max_{ v_i^k \in R} v_i^k| v_1^1,v_1^2,...,v_n^1,v_n^2\Big]
\end{equation}
In the last step, we have substituted the expression for expected maximum that we derived in \hyperref[eq:9]{(9)}. Thus we have the desired $2$-approximation. \hfill$\qed$

\subsection{Smoothing}
\label{sec:smoothing}
We mentioned that it is safe to assume that there is a strict ordering over the values. Otherwise, the gambler performs the following procedure. For all sample or real values $v_i$ that she sees, she privately samples $u_i \sim Unif(0,1)$ and represents the value as a a pair $(v_i, u_i)$. Whenever she encounters $v_i = v_j$, she compares $u_i$ to $u_j$ to determine which one is "larger". This is a standard smoothing argument which allows us to produce a strict ordering over the values with no performance loss.

\subsection{Adversaries}
Note that in our analysis of the algorithm, we assumed pessimistically that the algorithm accepts the smallest value which exceeds the threshold and proved the approximation guarantee under this assumption. This implies that the algorithm holds for a stronger adversary than usual. Using adversary types as defined in \cite{OCRS}, the standard prophet inequality has an \textit{offline adversary} who is only able to order the random variables before the game begins. In comparison, the \textit{almighty adversary} knows all of the random realizations before the game begins and can choose a different adversarial order for each set of realized values, and the worst-case ordering for our problem is smallest to largest. Thus we see that our algorithm maintains its guarantee against the almighty adversary as well.

\subsection{Alternative Algorithms}
We would like to note that the standard algorithm for the case in which we know the full distributions sets a fixed threshold of $\frac{1}{2}\mathbb{E}[\max_i X_i]$, rather than $\mathbb{E}[\max_i X_i]$, which would be the natural analogue of this single sample algorithm. For our single sample problem, it is interesting to note that for any $c \neq 1$, setting $c\max_i{v_i}$ as a fixed threshold does not guarantee any constant approximation.  

For $c > 1$, this statement is clear, as we can simply consider the case when a single distribution $X_1$ always has value 1 -- the algorithm will never select anything. For $c < 1$, consider $X_1$ which always has the value 1, and $X_2$ which has value $2^{n}$ with probability $\frac{1}{n}$ and $0$ otherwise. The expected maximum is $\frac{2^n}{n} + O(1) $, but the algorithm will only receive the large reward when the large reward comes in the sample, setting a high threshold, and it also comes in the real run. Therefore the expectation of the algorithm is $\frac{2^n}{n^2} + O(1)$, and we lose a factor of $n$.

\section{Mechanism Design}

This algorithm leads to a prior-independent order-oblivious posted price mechanism (OPM) for selling a single item to $n$ bidders with welfare guarantee of $\frac{1}{2}$ times the optimum. The reduction is straightforward and described in Azar et al. \cite{azar}, while OPMs were first introduced by Chawla et al. \cite{CHMS}. Although it is interesting that we can guarantee a 2-approximation to optimal welfare with a posted price, the VCG mechanism reduces to a second price auction for a single item and guarantees full welfare. Therefore we might like to study whether this leads to any revenue guarantees.

Borrowing Corollary 1 from \cite{azar}, we may derive a revenue bound for the case when the distributions are i.i.d. and regular.

\begin{corollary}
Let $\mathbb{M}$ be a comparison-based single-dimensional mechanism that guarantees an $\alpha$ approximation to welfare for all product distributions. Then when $X_i$ are i.i.d. and regular, $\mathbb{M}$ combined with lazy sample reserves guarantees an $\frac{\alpha}{2}$ approximation to revenue and welfare.
\end{corollary}
The condition \textit{comparison-based} requires that our algorithm only be based on the relative ordering of all of the values, rather than their magnitudes. We can see from our analysis that this condition does indeed hold. As described in Dhangwatnotai et al. \cite{dhangwatnotai}, the technique of sample reserves simply refers to setting a reserve price for each bidder that was equal to the random sample from their distribution. Setting reserves lazily refers to first running the mechanism and then removing bidders that failed to meet their reserve, and charging winners the maximum of the mechanism's price and the reserve price. However, note that any bidder whose value exceeds the maximum sample certainly exceeds their own sample, and the mechanism's price is equal to the maximum reserve price over all bidders, so reserves change nothing about our mechanism.

Thus, our algorithm can lead to a prior-independent single item mechanism which obtains $\frac{1}{4}$ of optimal revenue in the i.i.d. regular setting. Although \cite{dhangwatnotai} already showed a 2-approximation in this setting, our mechanism is simpler, using only a posted price.

In addition, we may also borrow Corollary 2 from \cite{azar} to prove a bound for MHR distributions.

\begin{corollary}
If $\mathbb{M}$ guarantees an $\alpha$ approximation to welfare and distributions are MHR, then $\mathbb{M}$ combined with lazy sample reserves guarantees an $\frac{\alpha}{2e}$ approximation to revenue and an $\frac{\alpha}{2}$ approximation to welfare.
\end{corollary}

Again, sample reserves do not change our mechanism, so our posted price guarantees a $\frac{1}{4e}$ approximation to optimal revenue for MHR distributions.

\section{Future Directions}
The most natural extension of our work would be to build on Azar, Kleinberg, and Weinberg's work \cite{azar2} to achieve a tighter competitive ratio in the $k$-choice prophet inequality with a single set of samples. We also have special interest in solving the single sample matroid prophet inequality, for which there is currently no known constant-competitive algorithm. The current best algorithm has a competitive ratio of $O(\log \log (\text{rank}))$, which combines the matroid secretary algorithm of Chakraborty and Lachish \cite{soda12} or Feldman, Svensson and Zenklusen \cite{secretary} with the reduction in \cite{azar2}. It is curious to note that the natural fusion of the idea from this paper and the technique from Kleinberg and Weinberg's paper \cite{matroid} does not lead to a constant-competitive algorithm. Thus, we are interested to see what innovations might be required to solve this problem.

\section{Acknowledgements}
This work was done under the guidance of Professor Matt Weinberg at Princeton University. We thank and acknowledge him for his crucial role in this research.

\end{document}